\documentclass{aa}

\begin{document}
\input{psfig.txt}
   \thesaurus{20(11.01.2, 11.19.1, 11.19.3, 11.17.2, 03.20.8, 04.19.1)
               }
   \title{Emission Line AGNs from the REX survey:}

   \subtitle{Results from optical spectroscopy\thanks{Partly based on
       observations collected at the European Southern Observatory, La
       Silla, Chile}}

   \author{A. Caccianiga\inst{1}, 
    T. Maccacaro\inst{2}, A. Wolter\inst{2}, R. Della Ceca\inst{2},
     I.M. Gioia\inst{3,4} }

   \offprints{A. Caccianiga, caccia@oal.ul.pt}

   \institute{\inst{1}Observat\'orio Astron\'omico de Lisboa, Tapada da Ajuda, 1349-018 Lisboa, Portugal
     \\
              \inst{2}Osservatorio Astronomico di Brera, via Brera 28, I-20121 Milano, Italy
              \\
              \inst{3}Institute for Astronomy, 2680 Woodlawn Drive, Honolulu, HI 96822
              \\
              \inst{4}Istituto di Radio Astronomia del CNR, via Gobetti 101, 40129 Bologna, Italy}
   \date{Received December 1, 1999; accepted March 16, 2000}

   \authorrunning{Caccianiga et al.}
   \maketitle

   \begin{abstract}
We present 71 Emission Line objects  selected from the REX
survey. Except for
3 of them, for which the presence of an active nucleus is
dubious, all these sources are Active Galactic Nuclei (QSOs,
Seyfert galaxies, emission line radiogalaxies).
In addition, we present the spectra of
other 19 AGNs included in a preliminary version of the 
REX catalog but not in the final one. 
The majority (80) of the 90 sources presented in this paper 
is newly discovered. Finally, we present the general  properties
in the radio and in the X-ray band of all the AGNs discovered
so far in the REX survey.

   \end{abstract}

%

\section{Introduction}
The REX survey is an effort aimed at the selection of a sizable,
statistically complete sample of Radio-Emitting X-ray sources (REXs).
A detailed description of this survey has been presented in
Caccianiga et al. (1999). In summary, the REX survey is the
result of a positional cross-correlation between the  
NRAO VLA Sky Survey (NVSS, Condon et al. 1998) at 1.4 GHz and
an X-ray catalog of about 17,000 serendipitous sources detected
in 1202 pointed ROSAT PSPC fields.
The flux density limit in the radio band (at 1.4 GHz) is  5 mJy
while in the X-ray band the flux limits range from 
$\sim$3.5$\times$10$^{-14}$ erg s$^{-1}$
cm$^{-2}$ to $\sim$2$\times$10$^{-13}$ erg s$^{-1}$
cm$^{-2}$ (0.5--2.0 keV). The area covered at 
the highest flux limit is about 2200 deg$^2$.
The cross-correlation has produced
a catalog of $\sim$1600 REX sources.
The spectroscopic observation of the REXs is in progress and, to date,
the percentage of identifications is about 30\%. Among these sources,
232 Emission Line (EL) objects (mostly AGNs), 72 BL Lacs and 176 optically
``non-active'' galaxies have been found.
About 30\% of these objects are newly discovered.
A previous paper  (Wolter et al. 1998) reported
a small fraction of these new identifications; 
in this paper we present the spectra of 71 
EL objects not included in Wolter et al. (1998).
In addition, we present other 19 EL objects that were included in a preliminary 
version of the REX survey but not in the final catalog. 

The paper is organized as follows: in section 2 we present 
the optical observations, in section 3 we explain the criteria 
used for the classification of the EL sources, in section 4 we 
discuss the general radio and X-ray properties of the EL AGNs 
discovered so far in the REX survey.
Our conclusions are summarized in section 5. 
Throughout this paper we use H$_0$=50~km $s^{-1}$ 
Mpc$^{-1}$ and q$_0$=0.

\section{Spectroscopic Observations}
Several spectroscopic observations of REX sources 
have been carried out during the 
period 1995/1998 using the 88" telescope of the University of Hawaii
(UH) in Mauna Kea (5 observing runs), the 2.2m and 3.6m telescopes of ESO
in La Silla (1 run) and the UNAM 2.1m telescope 
in S. Pedro Martir (3 runs). The results coming from the first  
two  observing runs at UNAM have been presented in
Wolter et al. (1998) and will not be considered here. 

The instrumental configurations are summarized in
Table~\ref{setup}. 
In all cases, we have used a long-slit and low dispersion (from 3.7 \AA/pixel
to 13.2 \AA/pixel) grating that maximizes the wavelength coverage.
For the data reduction we have used the IRAF {\it longslit} package. 
The spectra have been wavelength calibrated using an He-Ar 
(UNAM, ESO) or a Hg-Cd-Zn (UH) reference spectrum. 
The photometric standard stars used for the relative flux calibration are:
Feige~34 (UNAM 96/12), HD 19445 (UH 96/01,
UH96/08), LTT377 (ESO 96/12), SAO098781 (UH 97/03), 
HD84937 (UH 98/02), PG0216+032 (UH 98/10).

\begin{table*}
\begin{center}
\begin{tabular}{l l l l}
Telescope/Instrument & Grism name (g/mm) & 
Dispersion & Observing Period \\
\ & \ & \AA/pixel & \\
\ & \ & \ &  \\
\hline
UH 88''+ WFGS & blue (400) & 4.2 & 1996 Jan 14--15 \\
UH 88''+ WFGS & green (420) & 3.7 & 1996 Aug 7--11 \\
UNAM 2.1m + BC & (300) & 3.9 & 1996 Dec 6--10 \\
ESO 2.2m + EFOSC2 & G1 (100) & 13.2 & 1996 Dec 11--12 \\
ESO 3.6m + EFOSC1 & b300 (300), r300 (300) & 6.3, 7.5  & 1996 Dec 9--10 \\
UH 88''+ WFGS & blue (400) & 4.2 & 1997 Mar 3--5 \\
UH 88''+ WFGS & blue (400) & 4.2 & 1998 Feb 26 -- Mar 1 \\
UH 88''+ WFGS & blue (400) & 4.2 & 1998 Oct 15--18 \\ 
\hline
\end{tabular}
\end{center}
\caption{Observing setup}
\label{setup}

\vspace*{0.3 cm}



\end{table*}


\begin{table*}
\small
\begin{center}
\begin{tabular}{r l r r l l r}
Name & NVSS Position (J2000) & $F_X^a$ & $S_{1.4}^b$ & Set-up$^c$ & Date & Exposure \\
\ & \ & \ & \ & \ & \ & Time$^d$ \\
\hline

1REXJ000513$-$2614.6 & 00 05 13.71 $-$26 14 37.2 & 1.36  &   37.4 & 420  ; 2.3\arcsec  & 08/96   & 1200 \\ 
1REXJ001028+2047.8   & 00 10 28.80 +20 47 49.4   & 2.86  &  158.9 & 420  ; 2.3\arcsec  & 08/96 & 1200 \\ 
1REXJ002031$-$1510.8 & 00 20 31.06 $-$15 10 49.1 &1.93  &    16.6 & 420  ; 2.3\arcsec & 08/96   & 900 \\ 
1REXJ002841+0533.0   & 00 28 41.91 +05 33 04.0 & 1.56  &     7.1  & 420  ; 2.3\arcsec & 08/96   & 1200 \\ 
1REXJ004052$-$2902.2 & 00 40 52.42 $-$29 02 15.1 &3.95  &    49.5 & 400  ; 1.5\arcsec & 07/96   & 600 \\ 
1REXJ004413+0051.6   & 00 44 13.82 +00 51 40.7 & 0.71 &      40.5 & 420  ; 2.3\arcsec & 08/96   & 1800 \\ 
1REXJ005924+2703.5   & 00 59 24.40 +27 03 32.9 & 1.21  &    113.4 & 400  ; 1.5\arcsec &  10/98   &  420 \\
1REXJ011035$-$1648.5 & 01 10 35.13 $-$16 48 31.3 & 9.36  &   59.5 & 420  ; 2.3\arcsec & 08/96   & 2100 \\ 
1REXJ012210+0931.7   & 01 22 10.66 +09 31 44.9 &1.74 &        8.9 & 420  ; 2.3\arcsec & 08/96   & 900 \\ 
1REXJ012526+0856.5   & 01 25 26.80 +08 56 31.1 &1.58 &        9.9 & 420  ; 2.3\arcsec & 08/96   & 1800 \\ 
    J013707$-$2444.7 & 01 37 07.66 $-$24 44 47.8 & *0.53  &  21.2 & 420  ; 2.3\arcsec & 08/96   & 1800 \\ 
1REXJ014318+0228.3   & 01 43 18.58 +02 28 20.7 &0.45&        15.8 & 420  ; 2.3\arcsec & 08/96   & 1800 \\ 
1REXJ015232$-$1412.6 & 01 52 32.06 $-$14 12 38.2 &2.88  &   745.2 & 420  ; 2.3\arcsec & 08/96   & 900 \\ 
1REXJ020857$-$1003.2 & 02 08 57.04 $-$10 03 16.8 &0.42 &     27.0 & 400  ; 1.5\arcsec & 01/96   & 1800 \\ 
 1REXJ023556+1615.3  & 02 35 56.83 +16 15 23.9 & 5.04 &     129.9 & 400 ; 1.5\arcsec &    10/98   &  900 \\   
 1REXJ024613+1056.9  & 02 46 13.81 +10 56 56.7 &20.10 &      19.9 & 420  ; 2.3\arcsec & 08/96   & 900 \\
    J025057$-$1226.2 & 02 50 57.33 $-$12 26 15.8 & *2.30  &  14.6 & 400  ; 1.5\arcsec & 01/96   & 900 \\ 
    J025929+1925.7   & 02 59 29.65 +19 25 44.9 & *2.34  & 169.9 & 400  ; 1.5\arcsec & 01/96   & 1200  \\ 
1REXJ030459+0002.5   & 03 04 59.24 +00 02 33.6 & 3.13 &     124.9 & 400  ; 1.5\arcsec & 01/96   & 960 \\ 
1REXJ031958+0355.9   & 03 19 58.82 +03 55 56.4 & 7.20 &      55.7 & 400 ; 1.5\arcsec &    10/98   &  900 \\
    J033437$-$2559.5 & 03 34 37.63 $-$25 59 34.5 & *0.76  &  28.9 & 420  ; 2.3\arcsec & 08/96   & 1800 \\ 
1REXJ034026$-$2234.9 & 03 40 26.29 $-$22 34 54.2 & 1.14  &   96.0 & 420  ; 2.3\arcsec & 08/96   & 1200 \\ 
1REXJ041322+2343.5   & 04 13 22.50 +23 43 35.3 &  5.46 &     62.3 & 400 ; 1.5\arcsec &    10/98   &  900 \\
1REXJ041734$-$1154.5 & 04 17 34.91 $-$11 54 34.2 & 26.70 &   30.8 & 400 ; 1.5\arcsec &    10/98   &  1590 \\
     J053611+6027.3  & 05 36 11.23 +60 27 23.5 & *9.80 &     13.7 & 400 ; 1.5\arcsec & 01/96   & 900 \\
1REXJ061757+7816.1   & 06 17 57.03 +78 16 09.0 & 1.02 &     156.9 & 400 ; 1.5\arcsec &    02/98    & 1800 \\
1REXJ065154+6955.4   & 06 51 54.56 +69 55 26.4 & 0.97 &    276.3 & 400 ; 1.5\arcsec &    02/98    & 1800 \\   
     J071635+7108.6  & 07 16 35.45 +71 08 38.7 & *0.69 &    14.6 & 400  ; 1.5\arcsec & 01/96   & 2400 \\
 1REXJ071859+7124.3  & 07 18 59.61 +71 24 18.0 & 0.75 &    182.7 & 400  ; 1.5\arcsec & 01/96   & 1200 \\
 1REXJ073125+6718.7  & 07 31 25.55 +67 18 47.4 & 3.67 &     56.7 & 400 ; 1.5\arcsec &    02/98    & 1200 \\   
     J080017+3702.9  & 08 00 17.49 +37 02 59.8 & 1.81 &     30.9 & 400  ; 1.5\arcsec &03/97   & 900 \\ 
1REXJ081108+4533.8   & 08 11 08.81 +45 33 49.4 & 1.41 &     81.0 & 400 ; 1.5\arcsec &    02/98  &  1200 \\
1REXJ082656+6542.5   & 08 26 56.86 +65 42 31.9 & 0.72 &     34.4 & 400 ; 1.5\arcsec &    02/98    &  1800 \\
1REXJ082733+2637.2   & 08 27 33.76 +26 37 16.5 & 0.56 &    102.4 & 400 ; 1.5\arcsec &    02/98    &  1800 \\
1REXJ085211+7627.3   & 08 52 11.88 +76 27 18.2 & 1.62 &    191.9 & 400  ; 1.5\arcsec & 01/96   & 900 \\

\hline
\end{tabular}
\end{center}

\caption{Journal of observations carried out at the UH 88''. 
(continued on the next page)}

\end{table*}
\newpage

\begin{table*}
\small
\begin{center}
\begin{tabular}{r l r r l l r}
Name & NVSS Position (J2000) & $F_X^a$ & $S_{1.4}^b$ & Set-up$^c$ & Date & Exposure \\
\ & \ & \ & \ & \ & \ & Time$^d$ \\
\hline
 1REXJ092655$-$2345.4  &  09 26 55.95 $-$23 45 24.0 &   2.39 &   83.6 & 400 ; 1.5\arcsec &  02/98  & 1800 \\
     J095701+3207.0    &  09 57 01.54 +32 07 04.7 & *0.85  &     69.5 & 400  ; 1.5\arcsec & 01/96   & 1200 \\
 1REXJ102106+4523.4    &  10 21 06.01 +45 23 28.5 & 5.27  &     131.2 & 400  ; 1.5\arcsec &03/97   & 900 \\ 
 1REXJ102556+1253.8    &  10 25 56.33 +12 53 49.0 & 2.67 &      539.4 & 400 ; 1.5\arcsec &  02/98  & 1500 \\
 1REXJ103035+5132.5    &  10 30 35.18 +51 32 32.9 & 7.02 & 185.4 & 400 ; 1.5\arcsec &    02/98  & 1800 \\   
 1REXJ103206$-$1400.3  &  10 32 06.28 $-$14 00 19.9 & 1.32  & 196.1 & 400  ; 1.5\arcsec &03/97   & 1500 \\ 
 1REXJ121303+3247.6    &  12 13 03.81 +32 47 37.0 & 1.47  & 140.1 & 400  ; 1.5\arcsec & 01/96   & 1200 \\ 
 1REXJ121815+0744.4    &  12 18 15.55 +07 44 28.2 & 1.03 &  5.1 & 400 ; 1.5\arcsec &    02/98     & 900 \\   
 1REXJ123519+6853.6    &  12 35 19.22 +68 53 36.8 & 1.65 & 126.5 & 400 ; 1.5\arcsec &  02/98   & 1200 \\   
 1REXJ133714$-$1319.2  &  13 37 14.85 $-$13 19 16.8 & 2.80 & 114.5 & 400 ; 1.5\arcsec & 02/98  &  1500 \\   
 1REXJ134133+3532.8    &  13 41 33.14 +35 32 53.7 & 2.18  &  77.7 & 400  ; 1.5\arcsec &03/97   & 1800 \\ 
     J134252+4032.0    &  13 42 52.97 +40 32 01.5 & 4.05  &   152.2 & 400  ; 1.5\arcsec & 01/96   & 600 \\ 
 1REXJ134606+4859.6    &  13 46 06.12 +48 59 36.6 &  2.18 &  5.2 & 400 ; 1.5\arcsec &  02/98  &  1500 \\   
 1REXJ135409$-$0141.8  &  13 54 09.97 $-$01 41 50.4 & 1.76 &  39.1 & 400 ; 1.5\arcsec &    02/98 & 1500 \\   
 1REXJ140653+3433.6    &  14 06 53.86 +34 33 37.2 & 1.77  &  169.8 & 400  ; 1.5\arcsec &03/97 & 900 \\ 
 1REXJ141628+1242.2    &  14 16 28.64 +12 42 13.5 & 7.36 &  110.6 & 400 ; 1.5\arcsec & 02/98    & 2400 \\   
     J142744+3338.4    &  14 27 44.44 +33 38 28.6 & *0.83  &  16.3 & 400  ; 1.5\arcsec & 01/96   & 1500 \\ 
 1REXJ144544$-$2445.7  &  14 45 44.21 $-$24 45 42.1 & 4.57 & 210.5 & 400 ; 1.5\arcsec &  02/98  & 1800 \\   
 1REXJ152548+5828.8    &  15 25 48.20 +58 28 51.4 &  0.69 &  109.3 & 400 ; 1.5\arcsec &  02/98    & 1200 \\   
 1REXJ213248$-$0219.8  &  21 32 48.21 $-$02 19 50.5 &  7.80 & 29.1 & 400 ; 1.5\arcsec &    10/98  &  900 \\   
 1REXJ220451$-$1815.5  &  22 04 51.82 $-$18 15 35.4 & 10.40 & 38.5 & 400 ; 1.5\arcsec &    10/98  & 1800 \\   
 1REXJ223313+3405.0    &  22 33 13.05 +34 05 01.0 &  3.42 &  35.3 & 400 ; 1.5\arcsec &    10/98  &  900 \\   
 1REXJ230311$-$0859.3  &  23 03 11.03 $-$08 59 19.7 & 0.54 & 33.6 & 400 ; 1.5\arcsec &    10/98  &  480 \\ 
 1REXJ235029$-$2620.7  &  23 50 29.61 $-$26 20 46.1 &  8.86 &  11.1 & 400 ; 1.5\arcsec &   10/98   &  600 \\
 1REXJ235139$-$2605.0  &  23 51 39.37 $-$26 05 02.7 &  40.20 &  20.1 & 400 ; 1.5\arcsec &  10/98   &  900 \\ 
 

\hline
\end{tabular}
\end{center}

\setcounter{table}{1}
\caption{(Continued)}

\label{journal_mk}

\vspace*{0.3 cm}
$^a$ X-ray fluxes, corrected for Galactic absorption,
in the 0.5-2.0 keV band in units of 10$^{-13}$
erg s$^{-1}$ cm$^{-2}$; fluxes with an asterisk come from the 1RXP catalog
(see text for details)

$^b$ Radio flux densities at 1.4~GHz in mJy

$^c$ first column = Grism (400 = Grism 400~l/mm, 420 = Grism   
420~l/mm); second column = slit width

$^d$ Exposure time in seconds
\end{table*}

\begin{table*}
\small
\begin{center}
\begin{tabular}{r l r r l r}
Name & NVSS Position (J2000) & $F_X^a$ & $S_{1.4}^b$ & Set-up$^c$ & Exposure\\
\ & \ & \ & \ & \ & Time$^d$ \\
\hline
      J001341$-$3009.4 &  00 13 41.24 $-$30 09 26.6 & 0.43  &        223.0 & 2.2m; G1  & 1200 \\
 1REXJ005229$-$3743.8  &  00 52 29.01 $-$37 43 50.1 &0.52 &         30.2 & 3.6m; b300  & 600 \\ 
      J014716$-$0008.2 &  01 47 16.08 $-$00 08 17.7 &1.14  &          5.0 & 3.6m; b300  & 2400 \\ 
 1REXJ015429$-$1346.6  &  01 54 29.41 $-$13 46 36.4 &0.46 &         74.3 & 3.6m; b300  & 1200 \\ 
      J033953$-$2321.6 &  03 39 53.54 $-$23 21 36.7 & 0.33 &         36.6 & 3.6m; b300, r300  & 840, 300 \\
 1REXJ035348$-$1020.2  &  03 53 48.77 $-$10 20 14.3 & 0.70  &         19.1 & 2.2m; G1  & 1200 \\
      J040741$-$3102.7 &  04 07 41.85 $-$31 02 46.5 &  0.51  &         49.3 & 2.2m; G1  & 720 \\ 
 1REXJ041405$-$1224.2  &  04 14 05.96 $-$12 24 17.0 & 3.06  &         93.8 & 2.2m; G1  & 1200 \\ 
 1REXJ043851$-$2241.7  &  04 38 51.91 $-$22 41 44.3 & 0.83 &        217.0 & 2.2m; G1  & 720 \\ 
 1REXJ044105$-$1616.1  &  04 41 05.05 $-$16 16 07.1 & 1.26  &         45.1 & 3.6m; b300, r300  & 1200, 600 \\ 
 1REXJ044910$-$2015.4  &  04 49 10.33 $-$20 15 24.7 & 1.30  &         10.7 & 3.6m; b300  & 1500 \\ 
 1REXJ045756$-$2237.7  &  04 57 56.98 $-$22 37 45.8 & 2.70  &         11.5 & 2.2m; G1  & 1200 \\ 
 1REXJ053628$-$3401.1  &  05 36 28.46 $-$34 01 11.0 &  7.33  &        652.6 & 2.2m; G1  & 900 \\ 
 1REXJ054129$-$3427.7  &  05 41 29.75 $-$34 27 43.1 & 0.73  &         33.1 & 2.2m; G1  & 1080 \\ 
 1REXJ055722$-$1414.7  &  05 57 22.80 $-$14 14 42.4 & 1.74  &          5.5 & 3.6m; b300  & 1200 \\ 
 1REXJ062437$-$1824.1  &  06 24 37.43 $-$18 24 08.4 & 6.02  &          6.5 & 3.6m; b300  & 1260 \\ 
     J063114$-$2257.3  &  06 31 14.18 $-$22 57 19.8 & *1.64 &         19.8 & 3.6m; b300  & 300 \\ 
     J063155$-$2250.4  &  06 31 55.08 $-$22 50 24.2 & *1.95  &         77.1 & 3.6m; b300  & 1500 \\ 
 1REXJ085119+1358.4    &  08 51 19.81 +13 58 25.7 & 2.82  &         22.4 & 2.2m; G1  & 720 \\ 
 1REXJ085312+1358.8    &  08 53 12.21 +13 58 53.3 & 0.46 &         17.4 & 2.2m; G1  & 1500 \\ 
 1REXJ090015$-$2817.9  &  09 00 15.43 $-$28 17 58.6 & 2.61  &        512.1 & 2.2m; G1  & 1080 \\ 
     J091235$-$0956.3  &  09 12 35.74 $-$09 56 21.8 & 5.35 & 57.1 & 3.6m; b300  & 300 \\ 
 1REXJ105027$-$2816.0  &  10 50 27.31 $-$28 16 04.4 & 0.94  &         19.1 & 3.6m; b300  & 1260 \\ 
 1REXJ110744$-$3043.5  &  11 07 44.07 $-$30 43 35.6 & 1.44  &        351.5 & 3.6m; b300  & 600 \\ 

\hline

\end{tabular}
\end{center}

\caption{Journal of observations carried out at the ESO telescopes.}
\label{journal_eso}

\vspace*{0.3 cm}
$^a$ see Table~2

$^b$ see Table~2

$^c$ first column = telescope; second column = Grism. 
The slit width is always 1.5$\arcsec$

$^d$ see Table~2

\end{table*}

\begin{table*}
\small
\begin{center}
\begin{tabular}{r l r r c r}
Name & NVSS Position (J2000) & $F_X^a$ & $S_{1.4}^b$ & Slit width$^c$ & Exposure \\
\ & \ & \ & \ & \ & Time$^d$  \\
\hline
     J000640+2042.7   & 00 06 40.05   +20 42 45.0 & 0.50 & 80.5 & 1.6 & 4800 \\
 1REXJ022840$-$0935.2 & 02 28 40.81 $-$09 35 14.6 &  8.89 &  8.5  & 1.6 & 1800 \\
 1REXJ044754$-$0322.7 & 04 47 54.76 $-$03 22 43.2 & 10.80 & 87.3 & 1.6 & 900  \\
 1REXJ081041+0810.0   & 08 10 41.43   +08 10 00.5 & 3.33  &    198.7 & 1.6 & 3900 \\
     J091921+5048.9   & 09 19 21.77   +50 48 55.4 & *4.77 & 46.5 & 1.6 & 2220 \\
 1REXJ101238+5242.4   & 10 12 38.56   +52 42 25.2 & 3.74  &  8.1 & 2.4 & 3000 \\

\hline
\end{tabular}
\end{center}

\caption{Journal of observations carried out at the UNAM 2.1m.}

\label{journal_sp}

\vspace*{0.3 cm}
$^a$ see Table~2

$^b$ see Table~2

$^c$ Slit width in arcseconds
 
$^d$ see Table~2
\end{table*}

\begin{table*}
\small\
\begin{center}
\begin{tabular}{l l l l l l c c}
Name & Class & z  & C$^a$  & log$L_X^b$ & log$P_{1.4}^c$ & Emission Lines & Absorption Lines\\
\ & \ & \ & \ & \ & \ & \ & \\
\hline

1REXJ000513$-$2614.6 & B & 0.32 & F &  43.92 & 32.21$^*$ & $H\delta$,$H\gamma$,$H\beta$,[OIII],$H\alpha$ & \\ 
1REXJ001028+2047.8   & B & 0.60 & F &  44.88 & 33.46   & MgII,[OII],$H\delta$,$H\gamma$,$H\beta$,[OIII] &  \\ 
1REXJ002031$-$1510.8 & B & 0.59 & F &  44.69 & 32.26$^*$ & MgII,[OII],$H\gamma$,$H\beta$,[OIII] &  \\
1REXJ002841+0533.0   & B & 1.43 & T &  45.58 & 32.33$^*$ & MgII?  &   \\      
1REXJ004052$-$2902.2 & B & 0.265 & F & 44.19 & 32.15$^*$ & $H\delta$,$H\gamma$,$H\beta$,[OIII],$H\alpha$ & \\
1REXJ004413+0051.6   & B & 0.93 & F &  44.75 & 33.22$^*$ &   MgII,[NeV] &  \\     
1REXJ005229$-$3743.8$^{(1)}$ & B & 1.60 & F & 45.23 & 33.45$^*$ & CIV,CIII] &  \\  
1REXJ005924+2703.5   & N & 0.045 & F & 42.05 & 31.00 & $H\beta$,[OIII],$H\alpha$/[NII],[SII] & \\
1REXJ011035$-$1648.5$^{(2)}$ & B & 0.783 & F & 45.68 & 33.56 &  MgII,$H\delta$,$H\gamma$,$H\beta$ \\
1REXJ012210+0931.7   & B & 0.33 & F & 44.05 & 31.47$^*$ & [NeV],[OII],$H\gamma$,$H\beta$,[OIII],$H\alpha$ &\\
1REXJ012526+0856.5   & B & 0.895 & T & 45.06 & 32.25$^*$ & MgII? &  \\
1REXJ014318+0228.3   & B & 0.83 & T &  44.43 &   32.52$^*$ &   MgII? &  \\  
1REXJ015232$-$1412.6 & B & 1.35 & F &  45.78 &   35.00 &    CIII],MgII,[OII]  &  \\
1REXJ015429$-$1346.6 & B & 1.01 & F &  44.65 &   33.54 &   CIII],MgII  &  \\ 
1REXJ020857$-$1003.2 & B & 1.74 & F &  45.22 &   33.49$^*$ &  CIV,CIII] &  \\
1REXJ022840$-$0935.2 & N & 0.068 & F &43.28 & 30.19$^*$ & $H\gamma$,$H\beta$,[OIII],[OI],$H\alpha$,[SII] & \\
1REXJ023556+1615.3   & B & 0.256 & F &  44.26 & 32.57 & $H\delta$,$H\gamma$,$H\beta$,[OIII],$H\alpha$ & \\
1REXJ024613+1056.9   & B & 0.357 & F &  45.19 & 31.84 & $H\delta$,$H\gamma$,FeII,$H\beta$,[OIII],$H\alpha$ \\
1REXJ030459+0002.5$^{(3)}$ & B & 0.564 & F &  44.86 & 33.33 & MgII,[NeV],[OII],$H\delta$,$H\beta$,[OIII] & \\
1REXJ031958+0355.9   & B & 0.813 & F & 45.61 & 33.23 & MgII,$H\gamma$ & \\ 
1REXJ034026$-$2234.9 & B & 1.68 & F & 45.67 &  34.41 & CIII],FeII,MgII &   \\
1REXJ035348$-$1020.2 & B & 0.87 & F &  44.67 & 32.67$^*$ &  CIII],FeII,MgII,$H\beta$  & \\
1REXJ041322+2343.5  &  B & 0.309 & F & 44.49 &  32.41 & FeII,$H\gamma$,$H\beta$,[OIII],$H\alpha$  &  \\
1REXJ041405$-$1224.2$^{(4)}$& B &0.570 &F & 44.87 & 33.27 & MgII,[NeV],$H\delta$,$H\gamma$,$H\beta$,[OIII]&\\
1REXJ041734$-$1154.5  & N & 0.44 & F & 45.53 & 32.38$^*$ & [OII] & CaII H\&K \\
1REXJ043851$-$2241.7  & B & 1.66 & F &  45.52 & 34.83 & CIV,CIII],MgII &     \\
1REXJ044105$-$1616.1  & B & 0.408 & F & 44.13 & 32.53$^*$ & MgII,[OII],[OIII],$H\alpha$ & CaII H\&K\\
1REXJ044754$-$0322.7$^{(5)}$  & B & 0.773 & F & 45.73 & 33.48 & FeII,MgII,FeII,[OII],$H\delta$,$H\gamma$ &\\
1REXJ044910$-$2015.4  & B & 1.04 &  F & 45.14 & 32.40$^*$ & CIII],MgII   &    \\              
1REXJ045756$-$2237.7  & B & 0.12 & F &  43.28 & 30.80$^*$ & $H\alpha$  & CaII H\&K,G,MgI \\
1REXJ053628$-$3401.1  & B & 0.684 & F & 45.43 & 34.15 & MgII,[NeV],$H\delta$,$H\gamma$,$H\beta$,[OIII] & \\
1REXJ054129$-$3427.7  & B & 1.60 & F &  45.38 & 33.40$^*$ & SiIV/OIV],CIV,CIII],MgI &   \\                    
1REXJ055722$-$1414.7  & B & 0.358 & F & 44.13 & 31.26$^*$ & MgII,[OII],[NeIII],$H\delta$,$H\gamma$,$H\beta$,[OIII] & \\

\hline
\end{tabular}
\end{center}
\caption{Emission Line AGNs identified in the REX survey. (continued on the 
next page)}

\label{results}
\end{table*}

\setcounter{table}{4}
\begin{table*}
\small 
\begin{center}
\begin{tabular}{l l l l l l c c}
Name & Class & z  & C$^a$  & log$L_X^b$ & log$P_{1.4}^c$ & Emission Lines & Absorption Lines\\
\ & \ & \ & \ & \ & \ & \ & \\
\hline
1REXJ061757+7816.1   & B & 1.43 & F & 45.39 & 34.34 & CIII],MgII & \\
1REXJ062437$-$1824.1 & B & 1.89 & F & 46.48 & 32.40$^*$ & CIV,CIII] & \\  
1REXJ065154+6955.4   & B & 1.36 & F & 45.35 & 34.64 & CIII],MgII & \\   
1REXJ071859+7124.3$^{(6)}$ & B & 1.408 & F &  45.28 & 34.60 &  CIII], MgII & \\            
1REXJ073125+6718.7  & B & 0.17 & F & 43.74 &  31.78 & $H\gamma$,$H\beta$,[OIII],$H\alpha$ & CaII K   \\   
1REXJ081041+0810.0  & B & 0.391 & F & 44.51 & 33.20 & [OII],[NeIII],$H\gamma$,$H\beta$,[OIII] & \\
1REXJ081108+4533.8  & B & 1.02 & F & 45.15 &  33.50 & FeII,MgII,[OII],[NeIII],$H\delta$,$H\gamma$ & \\  
1REXJ082656+6542.5  & B & 0.956 & F & 44.79 &  33.41 &  MgII,$H\gamma$ &  \\                                
1REXJ082733+2637.2$^{(7)}$ & B & 0.69 & F & 44.34 &  33.52 & MgII,[OII],$H\gamma$,[OIII]& \\
1REXJ085120+1358.3  & B & 0.95 & F &  45.37 &   33.06 & CIII],MgII,[NeV],$H\delta$ &  \\
1REXJ085211+7627.3  & B & 1.127 &  F & 45.32 &   34.34 & CIII],MgII &  \\
1REXJ085312+1358.8  & B & 1.16 & F &  44.81 &   33.05$^*$ & CIII],MgII& \\
1REXJ090015$-$2817.9$^{(8)}$ & B & 0.89 & F & 45.29 & 34.46 & CIII],MgII,[NeV],[OII],[NeIII],$H\gamma$,$H\beta$,[OIII]& \\   
1REXJ092655$-$2345.4 & B & 1.17 & F & 45.54 & 33.72 & CIII]?,MgII,[NeV]?&  \\   
1REXJ101238+5242.4   & B & 0.129 & F &43.48 & 30.73$^*$ & [OII],$H\delta$,$H\gamma$,$H\beta$,[OIII],[OI],$H\alpha$,[SII]& MgI,NaID \\
1REXJ102106+4523.4$^{(9)}$ & B & 0.364 & F &  44.63 & 32.94 & $H\gamma$,$H\beta$,[OIII],$H\alpha$&  \\
1REXJ102556+1253.8$^{(10)}$ & B & 0.66 & T & 44.95 &  34.00 & MgII?,[OII]?,$H\beta$?& \\
1REXJ103035+5132.5  & B? & 0.518 & F & 45.12 & 33.40 & [OII],$H\gamma$,$H\beta$& CaII K \\
1REXJ103206$-$1400.3$^{(11)}$ & B & 1.059  & T &  45.17 & 34.00 & MgII?,[OIII]?&  \\
1REXJ105027$-$2816.0 & B & 0.41 & F & 44.01 & 32.11$^*$ & MgII,[NeV],[OII],$H\gamma$?,$H\beta$& CaII H\&K \\
1REXJ110744$-$3043.5 & B & 0.74 & F &  44.81 & 34.02 &   MgII,[NeV],[NeIII] &  \\                           
1REXJ121303+3247.6   & B & 2.507 &  F & 46.26 &   35.10 & $Ly\alpha$,CIV,CIII]&  \\        
1REXJ121815+0744.4   & N & 0.155 & F & 43.10 & 30.62$^*$ &[OIII],$H\alpha$,[SII] &   \\
1REXJ123519+6853.6   & B & 1.14 & F & 45.35 & 34.05 & CIII]?,MgII &  \\
1REXJ133714$-$1319.2 & B & 3.47 & F & 46.79 &  35.24 & $Ly\alpha$,OI,SiIV/OIV],CIV,CIII]&  \\
1REXJ134133+3532.8   & B & 0.783 & F & 45.05 &   33.36 & MgII,[OII],$H\gamma$,$H\beta$,[OIII]&   \\   
1REXJ134606+4859.6   & B & 0.187 & F & 43.60 & 30.82$^*$ & $H\alpha$& CaII H\&K \\
1REXJ135409$-$0141.8 & B & 0.98 & F & 45.20 &  33.25$^*$ & MgII,[OII]&  \\  
1REXJ140653+3433.6   & B &   2.56 & F &  46.28 &   34.76 &     $Ly\alpha$,SiIV/OIV],CIV,CIII]&    \\
1REXJ141628+1242.2   & B & 0.33 &  F & 44.68 & 32.74 & $H\gamma$,$H\beta$,[OIII],$H\alpha$& \\
1REXJ144544$-$2445.7 & B & 0.317 & F & 44.43 & 32.95 & $H\delta$,$H\gamma$,$H\beta$,[OIII],$H\alpha$& \\
1REXJ152548+5828.8   & B?& 0.309 & F & 43.59 & 32.70 & [OIII],$H\alpha$& CaII H\&K,G \\
1REXJ213248$-$0219.8 & B & 0.103 & F & 43.60 &  31.11$^*$&[NeIII]?,$H\delta$,$H\gamma$,$H\beta$,[OIII],$H\alpha$& \\
1REXJ220451$-$1815.5 & N & 0.21 & T & 44.39 & 31.86$^*$ & [OII]?,[OIII]?& \\
1REXJ223313+3405.0   & B & 0.63 & F & 45.01 & 32.76 & MgII,[NeV],[OII],[NeIII],$H\gamma$,$H\beta$,[OIII]& \\
1REXJ230311$-$0859.3$^{(12)}$ & N? & 0.024 & F & 41.14 &  29.92$^*$ &  $H\alpha$ & MgI,NaID  \\
1REXJ235029$-$2620.7 & B & 0.217 & F & 44.35 & 31.26$^*$ &[OII],[NeIII],$H\delta$,$H\beta$,[OIII],$H\alpha$&\\
1REXJ235139$-$2605.0 & N & 0.233 & F & 45.08 & 31.62$^*$ & [OII],$H\beta$,[OIII],[OI],$H\alpha$,[SII] & \\


\hline
\end{tabular}
\end{center}
\caption{(Continued)}


\vspace*{0.3 cm}
$^{a}$ redshift confidence: F=firm, T=tentative 

$^{b}$ De-absorbed (Galactic)
X-ray luminosity (0.5-2.0~keV band) in erg s$^{-1}$

$^{c}$ Monochromatic radio luminosity at 1.4~GHz in erg s$^{-1}$
Hz$^{-1}$. An asterisk indicates that the luminosity has been computed assuming a lower limit on the 
radio spectral index (see text for details);

\vspace*{0.3 cm}
Notes:

$^{(1)}$ Also in Iovino et al. (1996) (z=2.25 based on low dispersion prism observations)

$^{(2)}$ Also in Perlman et al. (1998) (z=0.78)

$^{(3)}$ Also in Perlman et al. (1998) (z=0.563)

$^{(4)}$ Also in Perlman et al. (1998) (z=0.569)

$^{(5)}$ Also in Perlman et al. (1998) (z=0.774)

$^{(6)}$ Also in Puchnarewicz et al. (1997) (z=1.419)

$^{(7)}$ Also in Puchnarewicz et al. (1997) (z=0.692)

$^{(8)}$ Also in Perlman et al. (1998) (z=0.894)

$^{(9)}$ Also in Laurent-Muehleisen et al. (1998) (z=0.364)

$^{(10)}$ Also in Perlman et al. (1998) (z=0.663)

$^{(11)}$ Also in Perlman et al. (1998) (z=1.039)

$^{(12)}$ Also in Da Costa et al. (1998) (z=0.02413)

\end{table*}

\begin{table*}
\small 
\begin{center}
\begin{tabular}{l l l l l l c c}
Name & Class & z  & C$^a$ &log$L_X^b$ &log$P_{1.4}^c$ & Emission Lines & Absorption Lines\\
\ & \ & \ & \ & \ & \ & \ & \\
\hline
J000640+2042.7    & B & 1.00 & T & 44.70 & 33.78 & MgII? & \\
J001341$-$3009.4  & B & 1.11 & F & 44.76 & 34.35 & CIII],MgII & \\  
J013707$-$2444.7$^{(1)}$ & B & 1.05 & T & 44.76 & 32.89$^*$ & MgII?  &   \\
J014716$-$0008.2  & N & 0.466 & F &  44.22 & 31.41$^*$  & [OII],[NeIII],H$\gamma$  &   \\  
J025057$-$1226.2  & B & 1.004 & T & 45.35 & 32.60$^*$ & MgII?   &      \\ 
J025929+1925.7    & B & 0.545 & F & 44.69 & 33.37 &  MgII,FeII,[NeV],$H\beta$ &    \\         
J033437$-$2559.5  & B & 1.16 & T & 45.03 & 33.19$^*$ & MgII? & \\   
J033953$-$2321.6  & B & 3.49 & F & 46.87 & 34.44$^*$ & OVI,$Ly\alpha$/NV,OI,SiIV/OIV],CIV,CIII]&  \\   
J040741$-$3102.7  & B & 1.40 & F  & 45.07 & 33.58$^*$ &   CIV, CIII], MgII & \\
J053611+6027.3    & B & 0.07 & F &  43.35 & 30.45$^*$ &  [OIII],[OI],$H\alpha$,[SII] &  \\  
J063114$-$2257.3  & B & 0.86 & T & 45.03 & 32.68$^*$ &   MgII?   &   \\  
J063155$-$2250.4  & B & 0.589 & F &  44.70 & 33.18$^*$ & MgII,[OII]& \\ 
J071635+7108.6    & B & 1.553 & F & 45.32 & 33.20$^*$ & CIV,OIII],CIII],MgII&  \\ 
J080017+3702.9    & B & 0.819 & F &  45.02 &   32.88 & MgII,FeII,[NeV],$H\gamma$,$H\beta$& \\
J091235$-$0956.3  & B & 0.361 & F & 44.63 &32.55$^*$ & MgII,[NeV],[NeIII],$H\gamma$,$H\beta$,[OIII]& \\
J091921+5048.9    & B & 0.915 & T & 45.58 & 33.45 & MgII?& \\
J095701+3207.0    & B & 0.334 & F & 43.76 & 32.62 & [OII],$H\delta$,$H\gamma$,$H\beta$,[OIII],$H\alpha$&  \\
J134252+4032.0$^{(2)}$ & B & 0.909 & T &  45.48 &   33.82 &  MgII? &  \\   
J142744+3338.4    & B & 1.237 & F & 45.14 & 33.07$^*$ & CIII],MgII & \\

\hline
\end{tabular}
\end{center}
\caption{Emission Line AGNs not included in the final version of the REX catalog.}

\vspace*{0.3 cm}
$^{a}$ see Table~5

$^{b}$ see Table~5

$^{c}$ see Table~5

Notes:

$^{(1)}$ Also in Lamer et al. (1997) (z=1.050)

$^{(2)}$ Also in Hook et al. (1998) (z=0.91)

\label{not_rex}
\end{table*}

In general, we have two exposures for each object, except for
few cases in which we have only one spectrum. 
The subtraction of the cosmic rays has been made manually, 
from the extracted spectrum.

On average, the seeing during the observing runs ranged from  
0.9\arcsec\ (August 1996, December 1996, February 1998) to 1.5\arcsec. 
We have consequently used, except for few cases, a slit of
1.5-1.6\arcsec\ to maximize the signal.

In Tables~\ref{journal_mk}, ~\ref{journal_eso} and ~\ref{journal_sp} 
we present the journals of the observing runs 
at  UH 88", ESO
2.2m/3.6m and UNAM 2.1m telescopes, respectively. 
For each object we report
name, NVSS position (J2000), X-ray flux corrected for Galactic 
absorption (in units of 10$^{-13}$
erg s$^{-1}$ cm$^{-2}$ in the 0.5-2.0 keV band), NVSS integrated flux
density at 1.4 GHz (in mJy), observing set-up, 
date of the observation and total exposure time.
The objects without the prefix ``REX" in the name do not
belong to the final REX catalog. 

The X-ray flux has been derived from the count-rate
using the value of Galactic N$_H$ at the source position
(Dickey \& Lockman 1990)
and assuming $\alpha_X$=1, which is the mean value
expected for our sources. 
For some of the sources that are
not in the final REX catalog, the X-ray flux has been computed
from the count-rate found in the 1RXP catalog 
(ROSAT NEWS n.32, 1994), converted from the  
0.1-2.4 keV to 0.5-2.0 keV band. These X-ray fluxes are
indicated with an asterisk in Tables~\ref{journal_mk}, ~\ref{journal_eso} 
and ~\ref{journal_sp}. The uncertainty on the
X-ray fluxes is about 20\%.
The typical rms of the radio (NVSS) maps is 0.45 mJy/beam\footnote{
For a detailed description of the 
uncertainties associated to the radio flux densities  
of the NVSS catalog,  see Condon et al. (1998)}.
We note that in the case of objects in cluster or in group
of galaxies the X-ray flux is the sum of the 
active nucleus plus the  extended thermal component. 
Given the available statistics and the 
spatial resolution of the PSPC, 
particularly poor in the external part
of the field, in most cases we are not able to distinguish the two
components. 
We have evidence, from the optical 
images, that 4 objects belong to a cluster/group of galaxies:
we believe that in these cases the X-ray luminosity is mainly due to the
diffuse intracluster gas (see Sect.~3). 

\section{The Emission Line objects}

In this paper we present newly identified Emission Line objects. 
We define Emission Line objects
those sources for which at least one strong
(EW$\gg$5~\AA\ in the source rest frame) 
emission line is present in the spectrum.
This criterium excludes BL Lac objects and   
radiogalaxies without emission lines in their
spectrum.  
There is evidence (Vermeulen et al. 1995; Corbett et al. 1996;  
March\~a et al. 1996) that some ``true'' BL Lacs might show 
emission lines with an equivalent width slightly 
larger than the ``canonical" limit of 5~\AA, 
proposed by Stocke et al. (1991).
However, the majority of the objects presented here show
very strong emission lines well above the limit of 5\AA.

We further classify objects as Broad Emission Line (BEL) AGNs if at 
least one line has a FWHM$>$1000 km/s (in the source rest frame),
or Narrow Emission Line (NEL) object if all the observed lines have FWHM$<$ 1000 km/s.

This classification reflects the common division between ``type 1'' 
(i.e. Broad Line) and ``type 2'' (Narrow Line) AGNs 
often discussed in literature. The majority of the EL objects
presented here is ``type 1'' AGNs. 
In one case (1REXJ213248$-$0219.8), the FWHM of H$\beta$ and 
H$\alpha$ is $\sim$1300-1400 km/s, thus slightly higher than the limit used to 
classify the object as NEL AGN. This value of FWHM and the ratio between 
[OIII] and H$\beta$ (=1.4) suggests a classification as narrow-line 
Seyfert~1 galaxy (NLS1, [OIII]/H$\beta<$3 and FWHM$<$2000 km/s, e.g. 
Osterbrock \& Pogge 1985).
In the few
cases of Narrow Line objects, we have applied, when possible,
the diagnostic criteria described in Veilleux \& Osterbrock (1987)
to separate a starburst galaxy from an AGN or LINER. 
However it is worth noting that, as discussed 
in Veilleux \& Osterbrock (1987), the  separation between 
starburst galaxies and AGNs
is not very sharp and a mixture of both  phenomena could be present 
in the same object  (Hill et al. 1999; Goncalves et al. 1998).
Given the  quality  of our spectra,
we didn't attempt to estimate a correction  for reddening.
However, for the classification of these objects we have
used, as long as possible, the ratios between couples of 
emission lines like [OIII]$\lambda$5007/H$\beta$,
[NII]$\lambda$6583/H$\alpha$,
[SII]$\lambda$6716+$\lambda$6731/H$\alpha$ that are not very 
reddening sensitive. 

We briefly discuss here the 4 NEL objects for which we can apply
the Veilleux \& Osterbrock (1987) diagnostic criteria.
 
{\bf 1REXJ005924+2703.5}: the values of the 
ratios between [SII]$\lambda$6716+$\lambda$6731 and H$\alpha$ 
(Log([SII]/H$\alpha$) = --0.29) and between 
[OIII]$\lambda$5007 and H$\beta$ (Log([OIII]/H$\beta$) = 0.02) 
put this object in the zone intermediate
between  AGNs and HII regions. 

{\bf 1REXJ022840-0935.2}: the intensity ratio between [OIII]$\lambda$5007 
and H$\beta$ (Log([OIII]/H$\beta$) = 0.26) and between 
[NII]$\lambda$6583 and H$\alpha$ (Log([NII]/H$\alpha$) =
--0.9)  suggests
a classification as H~II region or starburst galaxy rather than
AGN for this object. 

{\bf 1REXJ121815+0744.4}: the H$\alpha$ is blended 
with [NII]$\lambda$6548 and it is difficult to measure its 
width. The FWHM is about 800-1000 km/s. 
The strong intensity of the [OIII]$\lambda$5007 line 
if compared with the H$\beta$ (Log([OIII]/H$\beta$) = 0.88)
and the ratio between [SII]$\lambda$6716+$\lambda$6731 and H$\alpha$ 
(Log([SII]/H$\alpha$) = --0.37) strongly suggest the presence of an AGN.

{\bf 1REXJ230311-0859.3}: the H$\alpha$ is blended 
with [NII]$\lambda$6583. The FWHM of the H$\alpha$ is about 1000~km/s
although  a proper de-blending is necessary for a more accurate measurement. 
The measured value of Log([N II]$\lambda$6583/H$\alpha$)$\sim$--0.15 
and the lack of  [SII]$\lambda$6716+$\lambda$6731 lines 
lead to a tentative classification of
this source as Narrow Emission Line galaxy (NELG). This object
belongs to a galaxy pair. 

For the remaining 4 NEL objects (1REXJ041734-1154.5, 1REXJ220451-1815.5,
1REXJ235139-2605.0, J014716-0008.2) we cannot apply the diagnostic
criteria described above. These objects typically show [OII]$\lambda$3727
and [OIII]$\lambda$5007 lines. 
All these 4 NEL objects  show an X-ray luminosity exceptionally
high for this class of source (L$_X>$10$^{44}$ erg s$^{-1}$).
One of these sources, 1REXJ235139-2605.0, 
is the cD galaxy of the  cluster A 2667 which is very
luminous in the X-ray band (2.1$\times$10$^{455}$ erg s$^{-1}$
in the 0.1-2.4 keV band, Rizza et al. 1998).  
In other two cases (1REXJ041734-1154.5, 1REXJ220451-1815.5) 
the CCD image taken at the
UH 88'' telescope reveals the presence of an overdensity
of galaxies around ($<$ 1$\arcmin$) the radio source thus 
suggesting that these objects belong to a cluster/group of galaxies. 
Finally, J014716-0008.2 is probably in a cluster  since 
at least two companions at the same redshift of the radio
source, have been found. In conclusion,
all the NEL objects presented in this paper 
with an X-ray luminosity greater than
10$^{44}$ erg s$^{-1}$  probably belong to a cluster/group of
galaxies. In these cases, the very high 
X-ray luminosity is probably originated from the intracluster gas and not
from the galaxy itself. Further X-ray observations with higher 
spatial resolution are required in order to measure the
true  luminosity of the object. 
On the contrary, all the isolated NEL objects have a low luminosity 
in the radio and in the X-ray band (L$_X<$2$\times$10$^{43}$ erg s$^{-1}$,
L$_R\leq$10$^{31}$ erg s$^{-1}$ Hz$^{-1}$).

The classification of the EL objects in Broad Line (B) and Narrow Line (N)
objects is reported in Table~\ref{results} (column 2) together with 
the measured redshift and the relative confidence (column 3 and 4), 
X-ray and radio luminosities (columns 5 and 6), the
observed features in emission (column 7) and in absorption (column 8).
The luminosities have been computed and K-corrected under the assumption of
power-law spectra (f$_{\nu}\propto\nu^{-\alpha}$).
In the radio band we have used the spectral index
($\alpha_R$) as computed between 1.4~GHz (NVSS) and 5~GHz 
if the source is included in the GB6 (Gregory \& Condon, 1991) or 
in the PMN (Griffith \& Wright, 1993) catalog. 
For three sources (1REXJ061757+7816.1, 1REXJ082656+6542.5, 
1REXJ085211+7627.3) not present in the GB6 and PMN catalogs, we have used the 
flux at 365~MHz derived from the WENSS survey (Rengelink et al. 1997)
to estimate the radio spectral index.
Finally, if no fluxes at frequencies different from 1.4~GHz are available,
we have used the lower limit on $\alpha$ 
computed taking into account the proper flux density limit at 5~GHz 
for the GB6 ($\sim$ 18~mJy) and PMN catalogs (from 40 to 72~mJy
depending on the declination). These monochromatic luminosities 
(indicated by an asterisk in Table~\ref{results}) 
should be considered as lower limits. The actual value
of the luminosity is close to these values for low-redshift objects 
(z$<$0.5) while it can change up to a factor of 10 for those sources
at z$>$1.

\setcounter{figure}{2}
\begin{figure}[htbp]
\centerline{
\psfig{figure=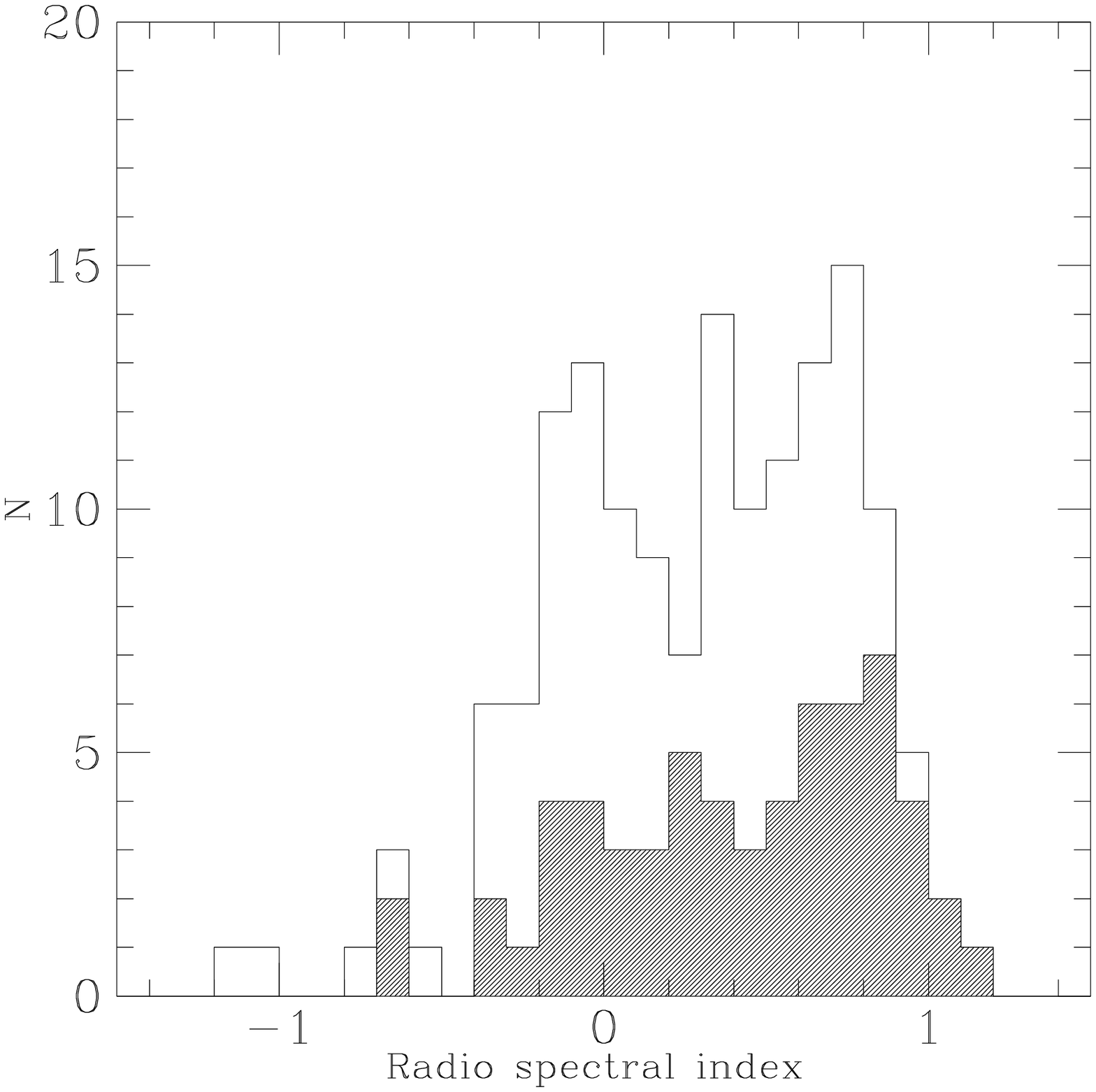,height=8cm,width=8cm}
}
\caption{Distribution of the radio spectral index between 1.4~GHz and 
5~GHz (f$\propto\nu^{-\alpha}$) for the AGN identified so far in the REX sample 
and for which a radio flux density at 5~GHz was available.
The shaded histogram represents the sources with a flux density at 1.4~GHz 
larger than 200~mJy and falling in the area of sky covered by
the GB6 catalog}
\label{alpha}
\end{figure}
In the X-ray band, the assumption of $\alpha_X$=1 is adequate  
to convert the count-rate into an X-ray flux (error less than 20\%) but 
it is not good enough for the
K-correction, in particular for high redshift objects. Thus, we 
have used the relationships found in
Brinkmann et al. (1997) that give the $\alpha_X$ as a function
of the redshift for a sample of X-ray selected radio-loud AGNs found in the
ROSAT All Sky Survey (RASS). In particular, we have used the appropriate
relationship for steep spectrum (SS, $\alpha_R>$0.5) and
flat spectrum (FS, $\alpha_R<$0.5) radio sources given in Brinkmann et al.
(1997) on the basis of the radio slope between 1.4~GHz and 5~GHz.
If a radio slope is not available, we have used the
relationship between $\alpha_X$ and z proper for the SS,
if the lower limit on $\alpha_R$ (based on the GB6 or PMN flux limits)
is larger than 0.5; otherwise, we have
used the formula for the FS.
The used values of $\alpha_X$ range from 0.8 to 1.15.

\begin{figure}[htbp]
\centerline{
\psfig{figure=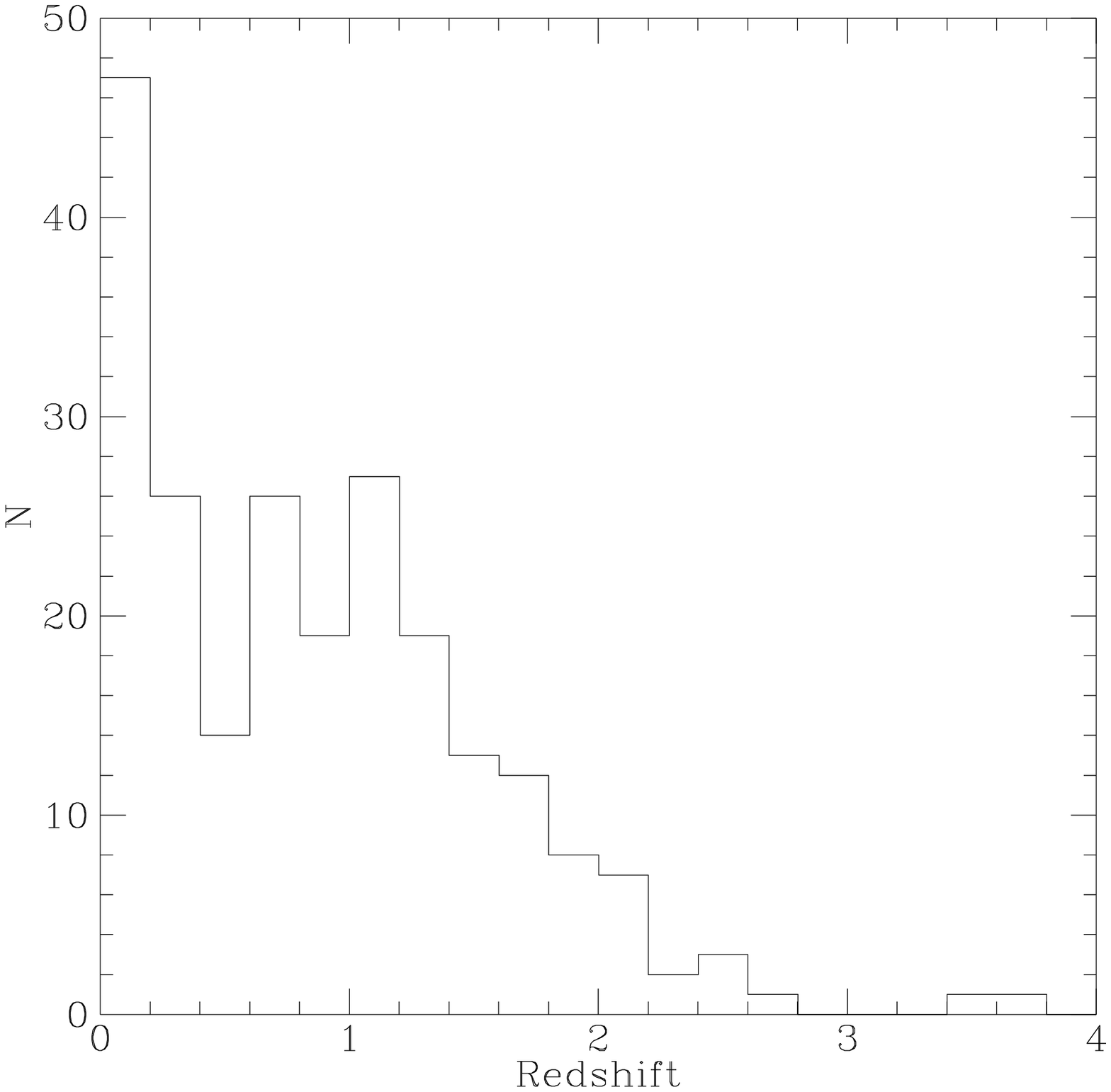,height=8cm,width=8cm}
}
\caption{Redshift distribution of the 226 AGNs found in the 
REX survey so far.}

\label{z}
\end{figure}

A separate Table (Table~\ref{not_rex})
contains the EL objects that are not included in the final version
of the REX catalog. These sources have been excluded because
they do not satisfy anymore the final selection criteria of the
REX survey described in Caccianiga et al. (1999).

We note that the spectra presented here
have been collected with the primary aim of identifying the 
sources and measuring the redshifts. The nominal error on the redshift 
is $\sim$0.001 (at 5000\AA) but in many cases, given the 
signal-to-noise ratio of the spectrum,
the error is larger (about 0.005 - 0.01).
When only one strong emission line is present in the spectrum, 
we have assumed, as usual, that this feature is the MgII$\lambda$2798. 
We have flagged as {\it tentative} the values of
redshift computed in these cases (a letter ``T'' after the value 
of redshift in  Table~\ref{results} and \ref{not_rex}).
We have also  flagged as {\it tentative} the redshift
of  1REXJ220451$-$1815.5 for which we have assumed that the
two observed features are [OII]$\lambda$3737 and
[OIII]$\lambda$5007 respectively.

All the sources presented here were unidentified 
when we observed them. 
During the preparation of this paper, 12 sources  
have been observed and identified independently
by other authors, in particular during the identification 
of other surveys that combine X-ray and radio data
(i.e. the DXRBS survey, Perlman et al. 1998 and the RGB survey,
Laurent-Muehleisen et al. 1998). Moreover, in two additional 
cases we have intentionally re-observed the object because
the redshift was based on poor quality data (i.e. low dispersion
prism observations, 1REXJ005229--3743.8) or we wanted to 
assess the presence of the $H_\alpha$ in emission 
(1REXJ230311--0859.3). Except for 1REXJ005229--3743.8,
for which we have found a redshift significantly different
from the one published (1.60 instead of 2.25), 
in all the other cases we confirm 
the published values of redshift, with discrepancies
below 3\% and only in one case of 6\%. A footnote in the source name
in Table~\ref{results} and \ref{not_rex} indicates that an optical
identification and a redshift for that object have been already 
published in literature.

In Fig.~\ref{sp_rex} and ~\ref{sp_nrex} we present the spectra of 
all the EL objects included
(Fig.~\ref{sp_rex}) and not included (Fig.~\ref{sp_nrex}) in 
the REX catalog.
The y axis (the flux) has arbitrary units.  

\section{Discussion}
Although the objects presented here do not represent a complete
sample, since the optical identification of the REX survey is still
in progress, it is interesting to investigate what kind of AGNs 
is selected by combining an X-ray and a radio survey. 
In order to increase the statistics, we have used for this 
analysis all the Emission Line AGNs discovered in the REX survey
so far. In particular, from the 232 EL objects available, we have 
selected  the 226 with a firm determination of redshift. 
The majority of the objects identified
from literature is composed by QSOs, i.e. type 1 objects.
For the low-redshift objects the situation is less clear 
in particular if we want to apply uniformly the same criteria used for the
sources observed by us. For this reason, in the following analysis
we will not distinguish between Narrow and Broad Emission Line
objects.

For all these sources we have computed the 
ratio ({\it r}) of the integrated to the peak NVSS flux densities,
resulting from the Gaussian fit to the image (see Condon et al. 1998
for details); {\it r} is a good indicator of
the compactness of the source at the survey resolution.
The majority (74\%) of the sources
has a {\it r} below 1.1, i.e. they are unresolved in
the NVSS maps. We note, however, that the
NVSS survey has been carried out with the VLA in the
less resolved configuration (DnC) and, thus, the corresponding
beam is quite large (FWHM=45\arcsec). This beam, 
at redshift of 0.9 (which is the mean z value of the 226 sources)
corresponds to a linear size
of about 470~kpc. Thus, even a typical lobe-dominated source at
this redshift could be unresolved in the NVSS.

In Fig.~\ref{alpha} we have reported the histogram of 
the spectral indices of the 151 sources detected at 5~GHz in the 
GB6 or PMN catalog. Ninety-four objects (62\%) are FS radio sources
($\alpha\leq$0.5) and 57 (38\%) are SS radio sources.
Since compact, flat spectrum
radio sources have a higher probability of being detected 
at 5~GHz (in the GB6 or PMN catalogs) we expect that
our distribution is biased against steep spectrum sources.
For this reason we have considered the 59 sources with a flux density 
at 1.4~GHz greater than 200~mJy and falling in the area of
sky covered by the GB6  catalog (which is deeper than PMN): 
all these sources should be detectable at 5~GHz even if their radio spectrum  
is very steep ($\alpha_R\sim$ 2). These sources, represented by the 
shaded area in Fig.~\ref{alpha}, are equally distributed
between FS (51\%) and SS (49\%) radio sources. 
Their average radio slope is 0.41.

\begin{figure}[htbp]
 \centerline{
\psfig{figure=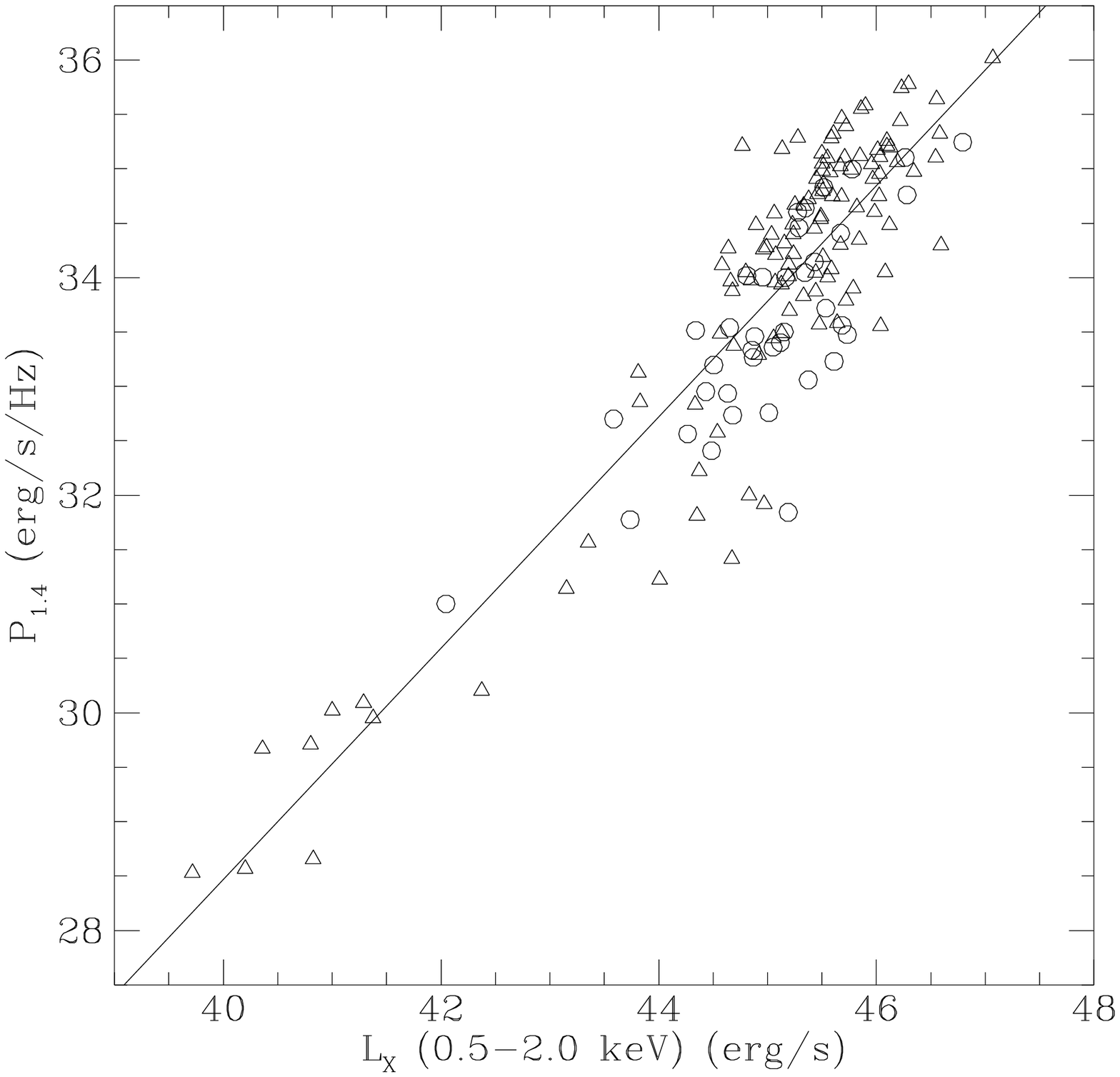,height=8cm,width=8cm}
}
\caption{Monochromatic radio luminosity versus X-ray luminosity for the 
AGNs belonging to the REX survey and for which we have a measured radio 
spectral index . The circles represent the objects
presented in this paper while the triangles
represent the objects previously identified. 
The {\it solid line} represents the best least squares fit to the data}

\label{lum}
\end{figure}

The redshift distribution of the 226 AGN is presented in Fig.~\ref{z}. 
The majority (96\%) of the sources have a redshift below 2. 
The mean value of z (=0.9) is the same found for the 
RL AGN in the EMSS (Della Ceca et al. 1994). 

In Fig.~\ref{lum} we present the monochromatic radio luminosities 
at 1.4~GHz 
versus the X-ray luminosities in the 0.5-2.0~keV band. 
We have considered here only the 151 AGNs for which we have 
computed the radio spectral index in order to reduce
the uncertainties on the determination of the K-corrected radio luminosity
which could be large for high redshift objects.
Fig.~\ref{lum} shows a rather strong correlation between the radio
and the X-ray luminosities. A least squares fit
to the data  gives L$_R \propto$L$_X^{1.06\pm0.04}$ 
(solid line in Fig.~\ref{lum}). 
This result is consistent (within 2$\sigma$) with a linear 
(L$_R \propto$ L$_X$) correlation between the two luminosities.
The Spearman Rank-Order correlation
coefficient gives a highly significant probability ($>$99.9\%) 
that the two luminosities are correlated. However, a spurious 
correlation between luminosities is often 
observed in flux-limited surveys, due to the artificial correlation
between luminosities and redshift. 
Since we do not see any evidence of a similar correlation between
the radio and X-ray fluxes, it is possible that what we 
observe in Fig.~\ref{lum} is simply the result of a selection 
effect. In order to exclude the effect of redshift in the 
analysis of the correlation bewteen the luminosities, we have used the partial 
correlation analysis described in Kendall \& Stuart (1979). 
According to this method, we have computed the correlation 
coefficient (r$_{xr.z}$) between L$_X$ and
L$_R$ excluding the dependence to the redshift in the following way:

\begin{center}
$r_{xr.z} = \frac{r_{xr} - r_{xz}r_{rz}}{\sqrt{1-r_{xz}^2}\sqrt{1-r_{rz}^2}}$
\end{center}

where $r_{xr}$, $r_{xz}$ and $r_{rz}$ are the Spearman Rank-Order 
correlation coefficients between L$_X$/L$_R$, L$_X$/z and L$_R$/z 
respectively. The analysis gives r$_{xr.z}$=0.09 corresponding to
a probability for the ``null hypothesis'' (the two luminosities 
are unrelated) of $\sim$25\%. Thus, once excluded the dependence to the 
redshift, our data do not show any strong evidence of a significant 
correlation between the two luminosities.

The distribution of X-ray luminosities is similar to that 
found for the RL AGNs in the EMSS (Della Ceca et al. 1994)
and it is peaked at L$_X$=5$\times$ 10$^{45}$ erg s$^{-1}$. 

\begin{figure}[htbp]
\centerline{
\psfig{figure=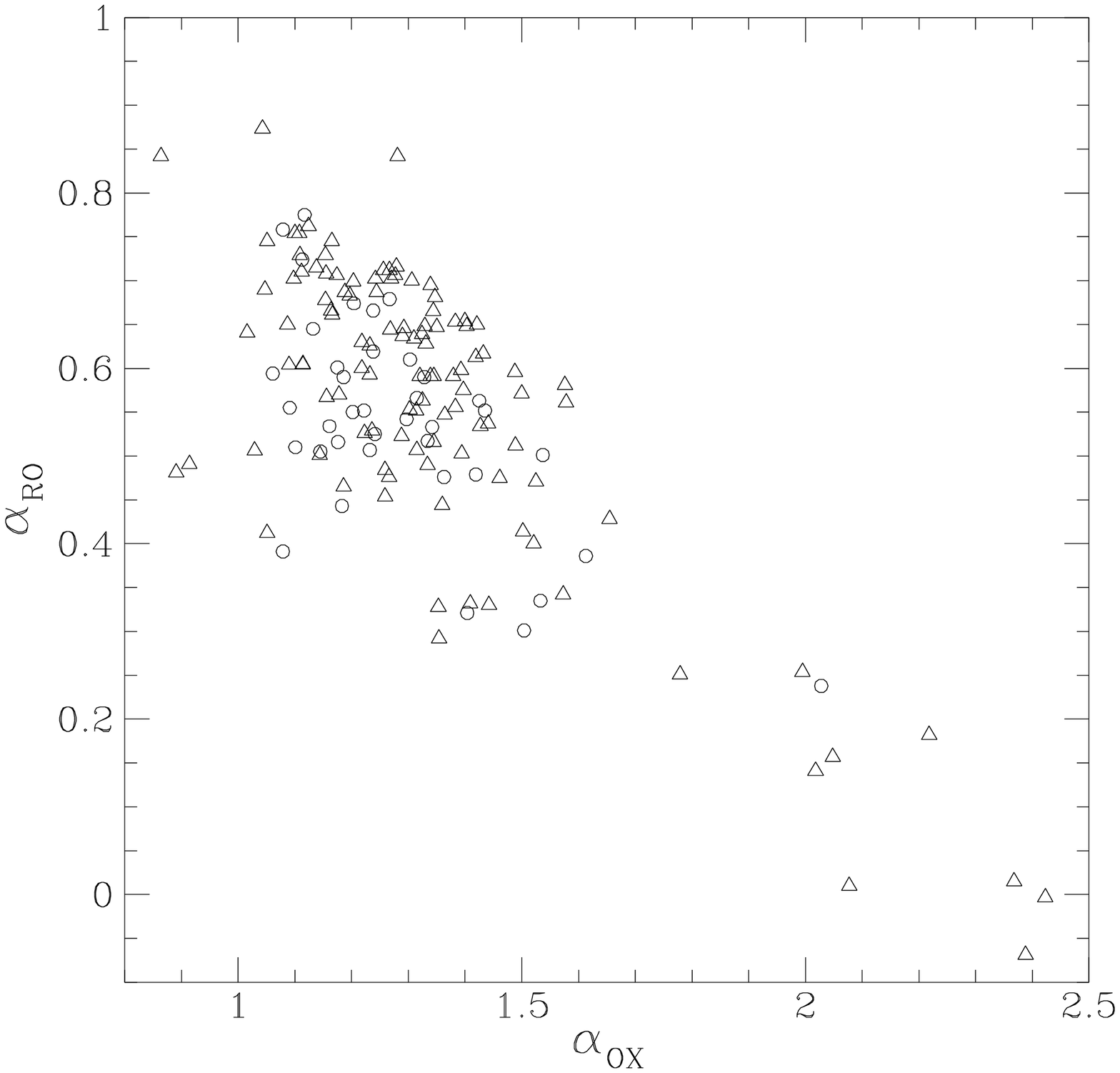,height=8cm,width=8cm}
}
\caption{Radio-optical ($\alpha_{RO}$) vs. the X-ray-optical
($\alpha_{OX}$) spectral indices of the AGN present in the 
REX survey. Symbols are the same as in Fig.~\ref{lum}}

\label{arox}
\end{figure}

In Fig.~\ref{arox} we report the Radio-optical ($\alpha_{RO}$) 
versus the X-ray-optical
($\alpha_{OX}$) spectral indices of the 151 AGNs presented
in Fig.~\ref{lum}. These indices 
are defined in the usual way (e.g. Stocke et al. 1991):

\begin{center}

$\alpha_{RO}=Log(S_{5GHz}/S_{2500\AA})/5.38$

$\alpha_{OX}=- Log(S_{2keV}/S_{2500\AA})/2.605$

\end{center} 

where $S_{5GHz}$, $S_{2500\AA}$ and $S_{2keV}$ 
are the K-corrected monochromatic flux densities at 5~GHz, 2500~\AA\ and
2~keV respectively. 
The optical flux densities have been derived
from the APM (Automatic Plate Measuring) O (blue) magnitude
assuming $\alpha_O$=1, while the
monochromatic fluxes at 2~keV have been computed from the X-ray fluxes
assuming $\alpha_X$=1. In the radio band we have used the
flux density at 5~GHz from GB6 or PMN.
About 88\% of the sources have an
$\alpha_{RO}\geq$0.35, which is the typical limit used to
define an object as ``radio-loud'' (RL, e.g. Della Ceca et al. 1994). 

\section{Conclusion}
We have reported the spectra and the main properties of 90 EL objects,
80 of which are newly discovered. 
Seventy-one objects belong to a well defined sample of
Radio-Emitting X-ray sources, the REX catalog  
(Caccianiga et al. 1999). The majority of the objects presented here
show broad (FWHM$>$1000 km/s) emission lines in the
optical spectrum while only 9\% are type 2 sources,
i.e. Narrow Emission line objects. Except for three sources, 
all the objects presented in this paper are AGNs
(QSOs, Seyfert galaxies, emission line radiogalaxies). 
The sample contains two high redshift (z=3.47 and 3.49) 
radio-loud AGNs.

We have then studied the general properties of the 226 AGNs discovered
so far in the REX sample, including sources identified
from literature.
We have analyzed their radio properties (compactness
and slope) as deduced from the NVSS data. On average,
we select a similar number of SS and FS radio sources
despite the fact that the selection has been done
by using a relatively low frequency (1.4~GHz) radio catalog.
At this frequency, we would have expected a dominance of 
SS AGNs. The large number of FS AGNs found in this
sample  is probably due to the presence of 
the X-ray constraint that favors the selection of compact
FS sources. On the other side, the presence
of a radio limit favors the selection of RL AGNs. 
In fact, the majority of these sources ($\sim$88\%) is radio loud  
while only a small percentage ($\sim$10\%, Della Ceca 
et al. 1994) of such objects is expected  in a purely X-ray 
selected survey. 

\begin{acknowledgements}
We are grateful to the staff at the  UH (Hawaii, USA), ESO (La Silla,
Chile)  and UNAM (San Pedro Martir, Mexico) telescopes for their support during the
observing runs. This research was partially  supported by the Italian Space 
Agency (ASI), by the European  Commission, TMR 
Programme, Research Network Contract ERBFMR
XCT96-0034 ``CERES", by the FCT under grant PRO15132/1999 and 
by the Italian Ministry for University and Research (MURST) under grant 
Cofin98-02-32. This research has made use of the NASA/IPAC 
Extragalactic Database (NED) which is operated
by the Jet Propulsion Laboratory, California Institute of Technology, 
under contract with the National Aeronautics and Space Administration.

\end{acknowledgements}

\newpage
\setcounter{figure}{0}
\begin{figure*}
\centerline{
}
\caption{
Optical spectra of the 71 EL AGNs included in the REX survey
\label{sp_rex}}
\end{figure*}

\newpage
\begin{figure*}
\centerline{
}
\caption{
Optical spectra of the 19 EL AGN not included in the REX survey. 
\label{sp_nrex}}
\end{figure*}

\end{document}